\documentclass[final,3p,times,twocolumn,authoryear]{elsarticle}

\usepackage{amssymb}
\usepackage{amsfonts}
\usepackage{booktabs}
\usepackage{subfigure}
\usepackage{amsmath}
\usepackage{amsthm}
\usepackage{multicol}
\usepackage{multirow}
\usepackage{algorithm}
\usepackage{algpseudocode}

\newtheorem{definition}{Definition}
\usepackage{color}

\journal{Expert Systems With Applications}

\begin{document}

\begin{frontmatter}



\title{Selective and Collaborative Influence Function \\for Efficient Recommendation Unlearning}

\author[zjucs]{Yuyuan Li}
\ead{11821022@zju.edu.cn}

\author[zjucs]{Chaochao Chen}
\ead{zjuccc@zju.edu.cn}

\author[zjucs]{Xiaolin Zheng\corref{cor1}}
\ead{xlzheng@zju.edu.cn}
\cortext[cor1]{Corresponding author}

\author[zjucs]{Yizhao Zhang}
\ead{22221337@zju.edu.cn}

\author[zjut]{Biao Gong}
\ead{a.biao.gong@gmail.com}

\author[oppo]{Jun Wang}
\ead{junwang.lu@gmail.com}

\address[zjucs]{College of Computer Science \& Technology, Zhejiang University, Hangzhou 310027, China}
\address[zjut]{Colledge of Computer Science \& Technology, Zhejiang University of Technology, Hangzhou 310023, China}
\address[oppo]{OPPO Research Institute, Shenzhen 518052, China}

\begin{abstract}
Recent regulations on the \textit{Right to be Forgotten} have greatly influenced the way of running a recommender system,
  because users now have the right to withdraw their private data.
  Besides simply deleting the target data in the database, unlearning the associated data lineage e.g., the learned personal features and preferences in the model, is also necessary for data withdrawal.
  Existing unlearning methods are mainly devised for generalized machine learning models in classification tasks.
  In this paper, we first identify two main disadvantages of directly applying existing unlearning methods in the context of recommendation, i.e., (i) \textit{unsatisfactory efficiency for large-scale recommendation models} and (ii) \textit{destruction of collaboration across users and items}.
  To tackle the above issues, we propose an extra-efficient recommendation unlearning method based on Selective and Collaborative Influence Function (SCIF).
  Our proposed method can (i) avoid any kind of retraining which is computationally prohibitive for large-scale systems, (ii) further enhance efficiency by selectively updating user embedding and (iii) preserve the collaboration across the remaining users and items.
  Furthermore, in order to evaluate the unlearning completeness, we define a Membership Inference Oracle (MIO), which can justify whether the unlearned data points were in the training set of the model, i.e., whether a data point was completely unlearned. 
  Extensive experiments on two benchmark datasets demonstrate that our proposed method can not only greatly enhance unlearning efficiency, but also achieve adequate unlearning completeness. 
  More importantly, our proposed method outperforms the state-of-the-art unlearning method regarding comprehensive recommendation metrics. 
  \end{abstract}




\end{frontmatter}


\section{Introduction}\label{sec:intro}

Recommender System (RS) has been extensively developed and deployed across various fields to extract valuable personalized information from collected user data~\citep{ji2020dual}.
Recently, there has been a surge in regulations aimed at preserving individual privacy. 
Notable among these regulations are the General Data Protection Regulation (GDPR)~\citep{voigt2017eu}, the California Consumer Privacy Act (CCPA)~\citep{2018ccpa}, and the Personal Information Protection and Electronic Documents Act (PIPEDA)~\citep{2019pipeda}. 
These regulations mandate companies to provide users with the option to withdraw their personal data that pertains to their privacy.
Apart from legal obligations, the need to enhance performance is another significant reason why companies opt to withdraw or remove data from their systems.
Existing study has shown that recommendation models, e.g., collaborative filtering, are highly sensitive to the training data~\citep{schafer2007collaborative}.
Often, these data can be tampered with inadvertently due to human errors or intentionally poisoned through malicious attacks~\citep{li2016data}.
By removing dirty data, the RS can enhance recommendation performance.

\begin{figure*}[t]
    \centering
    \includegraphics[width=\linewidth]{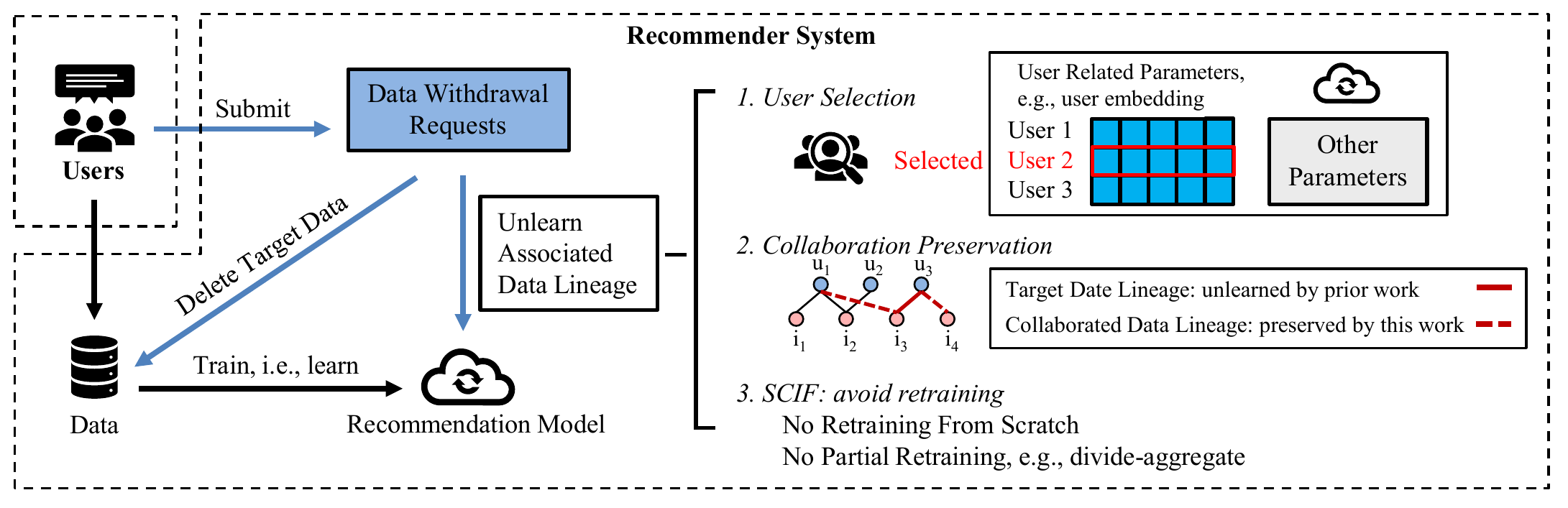}
    \caption{Overview of Selective and Collaborative Influence Function (SCIF) in recommendation unlearning.
    Note that in addition to users, the data withdrawal requests can be proactively initiated by the RS.}
    \label{fig:overview}
\end{figure*}

The mandatory withdrawal of data is significantly impacting the way RS is being operated.
In Figure~\ref{fig:overview}, the left part illustrates the workflows of a RS under this requirement, i.e., data withdrawal, including the original workflows (black arrows) and the add-on workflows (blue arrows).
As shown by the blue arrow in Figure~\ref{fig:overview}, in addition to deleting the target data in the database, it is also necessary for the RS to unlearn the associated data lineage in the learned recommendation models, i.e., \textit{recommendation unlearning}~\citep{chen2022recommendation}.
This stems from the fact that machine learning models, which are ubiquitously used in the RS, potentially memorize the training data~\citep{bourtoule2021machine,fredrikson2015model,carlini2019secret}. 
These memories have the potential to either compromise users' privacy or impact the accuracy of recommendations

In this paper, we focus on the recommendation unlearning problem.
Although several unlearning methods have been proposed in recent years, limitations still exist, especially in the context of recommendation.
To understand their deficiencies in the recommendation tasks, we first introduce three principles of unlearning, i.e., \textit{unlearning completeness (P1)}, \textit{unlearning efficiency (P2)}, and \textit{model utility (P3)}, then analyze the mechanisms of existing unlearning methods based on the above principles.
The most naive way of unlearning is retraining the model from scratch on the updated dataset after deletion.
Unfortunately, retraining from scratch costs tremendous computational overhead, which has driven the study of machine unlearning.
Considering the level of unlearning completeness, existing unlearning methods can be categorized into Exact Unlearning (EU, full completeness) and Approximate Unlearning (AU, partial completeness).
EU methods mainly focus on enhancing the retraining efficiency by either disassembling the model into reconfigurable components~\citep{ginart2019making,schelter2021hedgecut} or dividing the dataset into multiple shards~\citep{cao2015towards,bourtoule2021machine,yan2022arcane}.
However, the inherent mechanisms of existing EU methods lead to two key disadvantages for recommendation unlearning.
\begin{itemize}
    \item \textit{Unsatisfactory efficiency for the large-scale RS (against P2).} Although EU methods improve retraining efficiency to a certain extent, the continuous withdrawal requests from users can still cost prohibitive computational overhead for the large-scale RS in real-world scenarios. 
    More importantly, it is inconvenient to deploy existing methods in practice, since they \textit{change the original workflows} by splitting the model or dataset.
    \item \textit{Destruction of collaboration across users and items (against P3).} Due to the conflict between the inherent mechanisms of EU methods and the basic idea of recommendation, directly applying EU methods in recommendation potentially results in a significant performance decrease.
    This is because the basic assumption of most recommendation models is collaborative filtering, but existing EU methods impair the \textit{collaboration}, i.e., retrieving information through users-items connection, by splitting the model or dataset.
\end{itemize}

To further enhance efficiency and preserve collaboration, we follow the idea of AU and propose a novel recommendation unlearning method via Selective and Collaborative Influence Function (SCIF) in this paper.
Instead of retraining the model, AU methods perform reverse gradient operations on the learned model, which greatly improves unlearning efficiency.
Furthermore, the application of AU methods is particularly advantageous as they can be directly implemented on pre-existing, i.e., learned, models without disrupting their original workflow.
Nonetheless, there exist two notable drawbacks to the direct application of existing AU methods in recommendation tasks.
Firstly, the computation overhead of gradient operation, e.g., influence function, is still problematic for the large-scale RS (against P2).
Secondly, the impact of gradient operation regarding unlearning completeness (P1) and model utility (P3) is not clearly understood.
To tackle the above issues, our proposed SCIF approach first identifies the most pertinent user parameters in the model to reduce computational overhead. 
Following this, it considers the data of remaining users and items to restore collaboration. 
Once completed, SCIF then eliminates selective and collaborative influence on model parameters directly, without requiring retraining.
Obviously, completely unlearning the data lineage is a fundamental requirement of unlearning (P1), but there is no clear metric to evaluate the level of unlearning completeness in recommendation tasks yet.
Thus, we define the Membership Inference Oracle (MIO) to quantitatively analyze the unlearning completeness.
MIO is an authoritative membership inference attacker, which tells the probability of a given data point being in the model's training set.
If the target data is not in the training set of the unlearned model, we can affirm that the unlearned model has never seen the data, which means it achieves fully complete unlearning.
In this paper, we use neural networks to approximate MIO.
We summarize the main contributions of this paper as follows:
\begin{itemize}
    \item We propose an extra-efficient recommendation unlearning method via Selective and Collaborative Influence Function (SCIF), which not only enhances unlearning efficiency to a large extent, but also preserves collaboration to improve model utility. 
    \item We define the Membership Inference Oracle (MIO) as the unlearning completeness evaluation metric and build a neural network based approximated MIO to quantitatively analyze the unlearning completeness level of the proposed method.
    \item We conduct extensive experiments on two benchmark datasets. 
    The results show that our proposed method is capable of achieving not only satisfactory unlearning completeness but also delivering consistent enhancements in both unlearning efficiency and recommendation performance when compared to the state-of-the-art unlearning method.
\end{itemize}

\section{Related Work}

\subsection{Machine Unlearning}

Due to the fact that retaining from scratch is computationally prohibitive for large-scale machine learning models, a number of studies have been carried out to address the issue of inefficiency in the machine unlearning problem.
Based on the level of unlearning completeness, existing unlearning methods can be categorized into two approaches, i.e., exact unlearning and approximate unlearning.

\paragraph{Exact Unlearning (EU)} This approach aims to assure that the data lineage is completely unlearned from the model.
As retraining is an authoritative way to achieve completeness, this approach focuses on enhancing retraining efficiency.
The idea of existing EU methods is divide-aggregate, which means dividing the dataset or model into sub-components, training them separately, and aggregating them in the end~\citep{cao2015towards, ginart2019making, schelter2021hedgecut, bourtoule2021machine, yan2022arcane}.
With the help of the divide-aggregate framework, EU methods can limit the overhead of retraining to sub-components, and avoid retraining the model on the whole dataset, i.e., retraining from scratch.
EU methods inherently achieve unlearning completeness (P1), but suffer from a trade-off between unlearning efficiency (P2) and model utility (P3).
On the one hand, increasing the number of sub-components can improve unlearning efficiency. 
But doing so has a potential downside.
As the number of components grows, each component risks becoming a weak learner, which ultimately diminishes model utility.
On the other hand, limiting the number of sub-components can help preserve the model's overall utility. 
But doing so also curtails the efficiency of unlearning.

\paragraph{Approximated Unlearning (AU)} This approach aims to achieve real-time unlearning by avoiding any kind of retraining.
Existing AU methods estimate, i.e., approximate, the influence of target data and directly remove it through reverse gradient operations~\citep{sekhari2021remember, wu2022puma, mehta2022deep}. 
Estimating the influence of target data is mainly based on influence function~\citep{koh2017understanding, koh2019accuracy}.
Although AU methods can theoretically boost the unlearning efficiency, the computational overhead of influence estimation is still prohibitive for large-scale models.
The latest AU methods manage to accelerate influence estimation by approximation, i.e., approximate the approximation~\citep{wu2022puma, mehta2022deep}, which inevitably results in decreased accuracy of influence estimation.
As a comparison, our proposed method identifies the most pertinent parameters for computation, thereby reducing the computational overhead from a fundamental level.

Existing unlearning methods mainly focus on classification tasks while little attention has been paid to recommendation tasks.
Thus, as we mentioned in Section~\ref{sec:intro}, it is inappropriate to directly apply existing unlearning methods in recommendation tasks.

\subsection{Recommendation Unlearning}

Recommendation is a real-world scenario where unlearning is of great demand.
RecEraser was proposed to achieve unlearning in recommendation tasks~\citep{chen2022recommendation}.
Following EU's divide-aggregate framework, RecEraser groups similar data together.
This modification allows RecEraser to preserve collaborative information, which is important to the performance of personalized recommendation.
RecEraser also uses an attention-based aggregation to further enhance model utility.
However, RecEraser still suffers from the inherent disadvantages of EU methods, e.g., unsatisfactory unlearning efficiency, and limited collaboration preservation.

\subsection{Membership Inference}

Membership inference is a well-acknowledged method used to analyze information leakage from a trained model~\citep{yu2021does}.
Specifically, given a trained model (target) and a data point (query), membership inference determines whether this point was in the model's training dataset.
Membership inference attack against machine learning models was pioneered by~\cite{shokri2017membership}.
The main idea is regarding the membership inference problem as a binary classification task, and using machine learning classifiers to attack the target machine learning model.
To improve the performance of machine learning classifiers, i.e., attacker, \cite{shokri2017membership} uses shadow models, which simulate the behavior of target model, to generate sufficient training data for the attacker.
The following work has investigated various settings of shadow model training and presented several defence techniques ~\citep{salem2018ml,wu2020characterizing,yu2021does}.
However, current research has primarily concentrated on exploiting shallow classification models, leaving deep learning and regression models such as collaborative filtering largely unexplored.

\section{Preliminaries}

In this section, we first identify three principles of unlearning. Afterwards, we briefly introduce the notations of unlearning.

\subsection{Principles of Unlearning}\label{sec:goal}

We identify three principles that we consider as the pillars of achieving successful unlearning in recommendation tasks.
Similar objectives can also be found in~\cite{chen2022recommendation}.

\paragraph{P1: Unlearning Completeness} 
Completely unlearning the associated data lineage w.r.t. the target data is one of the most fundamental requirements of unlearning. 
Full completeness means totally eliminating the target user information learned by recommendation models and making it impossible to recover.
It is also reasonable to trade completeness for efficiency in real-world scenarios. 
We introduce an ideal metric, i.e., MIO, to evaluate the level of completeness in Section~\ref{sec:eva}. 

\paragraph{P2: Unlearning Efficiency}
Efficiency is another important principle of unlearning.
Due to the considerable computational overhead of practical recommendation models, including both time and space costs, retraining from scratch is prohibitive.

\paragraph{P3: Model Utility}
It is obvious that recommendation platforms do not want to experience a decline in recommendation performance after unlearning.
However, the fact is that unlearning too much data lineage will inevitably reduce the model utility, because unlearning is equivalent to reducing the amount of training data. 
Thus, an adequate unlearning method needs to generate an unlearned model that achieves comparable performance with the model retrained from scratch.

\subsection{Unlearning Notation}\label{sec:nota}

During the original workflows (black arrows in Figure~\ref{fig:overview}), the RS trains a model $\mathcal{M}(D)$ based on a dataset $D = \{d_u| u \in \{1, ..., N\}\}$ from $N$ users where $d_u$ represents the data of user $u$. 
During the unlearning workflows (blue arrows in Figure~\ref{fig:overview}), a user $u$ can submit any withdrawal request $E \subset d_u$ to unlearn the personal data.
Typically, $|E| \ll |D|$, where $|\cdot|$ denotes the number of samples.
%
%
An unlearning method $h$ maps the original model $\mathcal{M}(D)$ into the unlearned model $\mathcal{M}_{\lnot E}(D)$, which is denoted as $h: \mathcal{M}(D) \times E \mapsto \mathcal{M}_{\lnot E}(D)$.
Based on the affinity of the unlearned model and the model retrained from scratch, we define two types of unlearning as follows:
\begin{definition}[Strong Unlearning]\label{def:opt}
    We define that an unlearning method achieves strong unlearning, if the distribution of the unlearned model's parameters is identical to that of the model which is trained from scratch. Formally, 
    \begin{equation}
        h(\mathcal{M}(D), E) = \mathcal{M}_{\lnot E}(D) =_p \mathcal{M}(D/E),
    \end{equation}
    where $=_p$ denotes distributional equality w.r.t. model parameters.
\end{definition}

\begin{definition}[Weak Unlearning]
    We define that an unlearning method achieves weak unlearning, if the distribution of the unlearned model's outputs is identical to that of the model which is trained from scratch. Formally,
    \begin{equation}
        h(\mathcal{M}(D), E) = \mathcal{M}_{\lnot E}(D) =_o \mathcal{M}(D/E),
    \end{equation}
    where $=_o$ denotes distributional equality w.r.t. model outputs.
\end{definition}

Intuitively speaking, strong unlearning expects that the unlearned model has never seen the target data $E$.
In contrast, weak unlearning only expects that the unlearned model behaves as if it has never seen the data.
Regrettably, a weak unlearning model can still memorize the data lineage, leaving it vulnerable to various attacks, e.g., model inversion attack~\citep{fredrikson2015model}.
On the one hand, the majority of existing work~\citep{cao2015towards,bourtoule2021machine,chen2022recommendation,yan2022arcane} focuses on strong unlearning, which perfectly achieves P1 and P3.
On the other hand, the study of weak unlearning is of great value in practice~\citep{baumhauer2020machine}, since it is considerably more efficient than strong unlearning, and it is acceptable to sacrifice completeness for efficiency in some real-world scenarios.
In this paper, we aim to propose an efficient strong unlearning method in the context of recommendation.

\section{Recommendation Unlearning}

In this section, we first identify the inherent disadvantages of Exact Unlearning (EU) methods and the weakness of existing Approximate Unlearning (AU) methods. 
Then we propose an efficient recommendation unlearning method based on Selective and Collaborative Influence Function (SCIF).

\subsection{Disadvantages of Exact Unlearning}

\begin{figure}[t]
    \centering
    \includegraphics[width=0.7\linewidth]{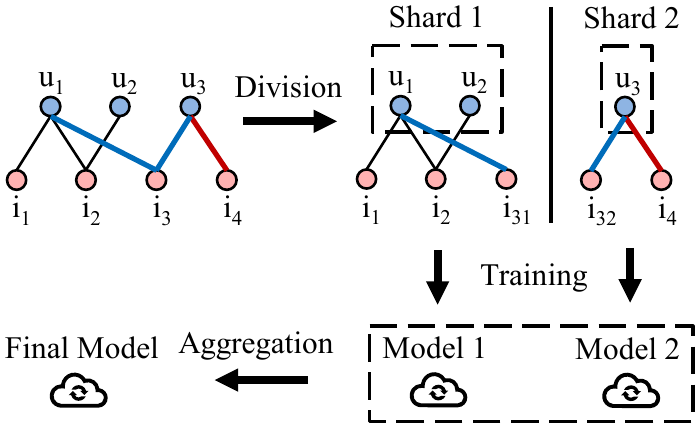}
    \caption{Illustration of divide-aggregate framework where $u$ and $i$ denote users and items respectively. The edges between users and items are ratings, and we assume rating $(u_3, i_4)$ is unlearned. Divide-aggregate framework trains individual models on the previously divided sub-components in isolation and then aggregates them. 
    }
    \label{fig:coll}
\end{figure}

Existing EU methods are mainly based on the divide-aggregate framework, which arises two inherent disadvantages in recommendation tasks, as we mentioned in Section~\ref{sec:intro}.
We now further describe these two disadvantages using the example in Figure~\ref{fig:coll}.

First of all, the efficiency of the divide-aggregate framework is unsatisfactory.
In Figure~\ref{fig:coll}, $u_3$ requires to unlearn rating $(u_3, i_4)$.
Instead of retraining the final model on the whole dataset, the divide-aggregate framework only needs to retrain \textit{Model 2} on \textit{Shard 2}.
Although it can enhance unlearning efficiency to some extent, it still requires partial retraining, e.g., \textit{shard 2} in this example, which can be computationally prohibitive in real-world scenarios. 
Note that over-increasing the number of sub-components can result in the issue of weak learners.
Moreover, the divide-aggregate framework cannot be applied to a pre-existing model.
In other words, if a RS wants to achieve unlearning, it has to follow the divide-aggregate learning workflow in advance.
Therefore, it is inconvenient for the pre-existing RS to deploy the divide-aggregate unlearning method in practice.

Secondly, disassembling the model or splitting the dataset is not trivial and can possibly affect model performance.
Note that a lot of machine learning tasks have the characteristics of association-sensitive. 
That is, the tasks rely on data association. 
Recommendation is a typical association-sensitive task, which relies on collaboration across users and items~\cite{shi2014collaborative}.
We can observe from Figure~\ref{fig:coll} that the divide-aggregate framework impairs collaboration during learning.
Before division, $u_1$ associates with $u_3$ through $i_3$.
After division, the connect is cut off, and each sub-component learns its own embedding of $i_3$ ($i_{31}$ and $i_{32}$).
Although aggregation can reunite all sub-components in the end, it cannot restore the collaboration impaired during division.

\subsection{Recommendation Unlearning with SCIF}

In order to overcome the inherent disadvantages of EU methods, we follow the idea of AU and propose an efficient recommendation unlearning method via Selective and Collaborative Influence Function (SCIF).
Compared with existing Influence Function (IF) based AU methods, SCIF incorporates two key components, i.e., user selection and collaboration preservation, to improve its unlearning performance in recommendation tasks.
Specifically, user selection is used to reduce the number of parameter updates by selecting the most pertinent user embedding.
collaboration preservation is used to enhance model utility by considering the collaborative data of the target data.

\subsubsection{Influence Function}

AU methods directly remove the influence of target data by reverse operation, avoiding any kind of retraining.
We first introduce IF which is used by AU methods to estimate influence in unlearning.
Let us start with a general machine learning task that is described by a dataset $D = \{z_1, ..., z_n\}$ and a loss function $\ell(z_i, \theta)$ where $\theta$ denotes model parameters.
We assume that the loss function is twice-differentiable and strictly convex.
Following the empirical risk minimization framework, the minimizer is given by
\begin{equation}
    \theta^* = \underset{\theta}{\operatorname{\arg\min}}\hspace{1mm}\sum_{i=1}^n \ell(z_i, \theta),
\end{equation}
where we omit the regularization term in the loss function for conciseness.

According to Definition~\ref{def:opt}, strong unlearning requests to generate an unlearned model that has never seen the target data.
Thus, unlearning a data point $z$ can be described as follow:
\begin{equation}
    \theta^*_{\lnot z} = \underset{\theta}{\operatorname{\arg\min}}\hspace{1mm}\sum_{i=1}^n \ell(z_i, \theta) - \ell(z, \theta),
\end{equation}
where $\theta^*_{\lnot z}$ is the parameter of unlearned model, i.e., retraining from scratch.

Koh and Liang~\cite{koh2017understanding} have studied the problem of model parameter change, i.e., influence on model parameters, when weighting a data point $z$ by $\epsilon$.
Formally, the $\epsilon$-weighted parameter is computed as
\begin{equation}
    \theta^*_{\epsilon, z} = \arg\min_{\theta}\sum_{i=1}^n \ell(z_i, \theta) + \epsilon\ell(z, \theta).
\end{equation}
According to~\cite{koh2017understanding}, the IF of $z$ is given by
\begin{equation}\label{equ:if}
    \mathcal{I}(z) := \frac{d\theta^*_{\epsilon, z}}{d\epsilon}\Big\vert_{\epsilon=0} = -H_{\theta^*}^{-1}\nabla_\theta \ell(z, \theta^*),
\end{equation}
where $\nabla_\theta\ell(z, \theta^*)$ is the gradient vector, and $H_{\theta^*} := \sum_{i=1}^n\nabla^2_\theta\ell(z, \theta^*)$ is the Hessian matrix and is positive definite by assumption.

\begin{figure}[t]
    \centering
    \includegraphics[width=0.95\linewidth]{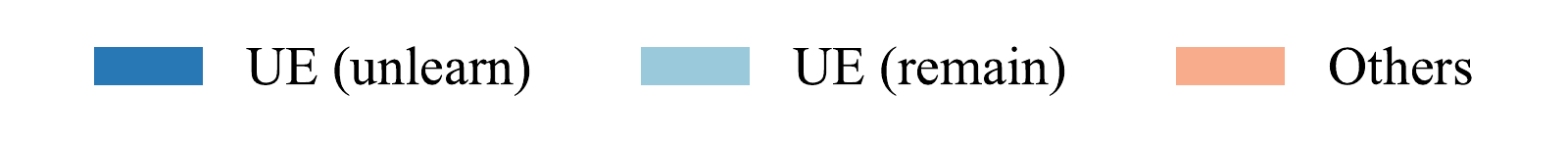}\\
    \subfigure[NMF]{
        \includegraphics[width=0.47\linewidth]{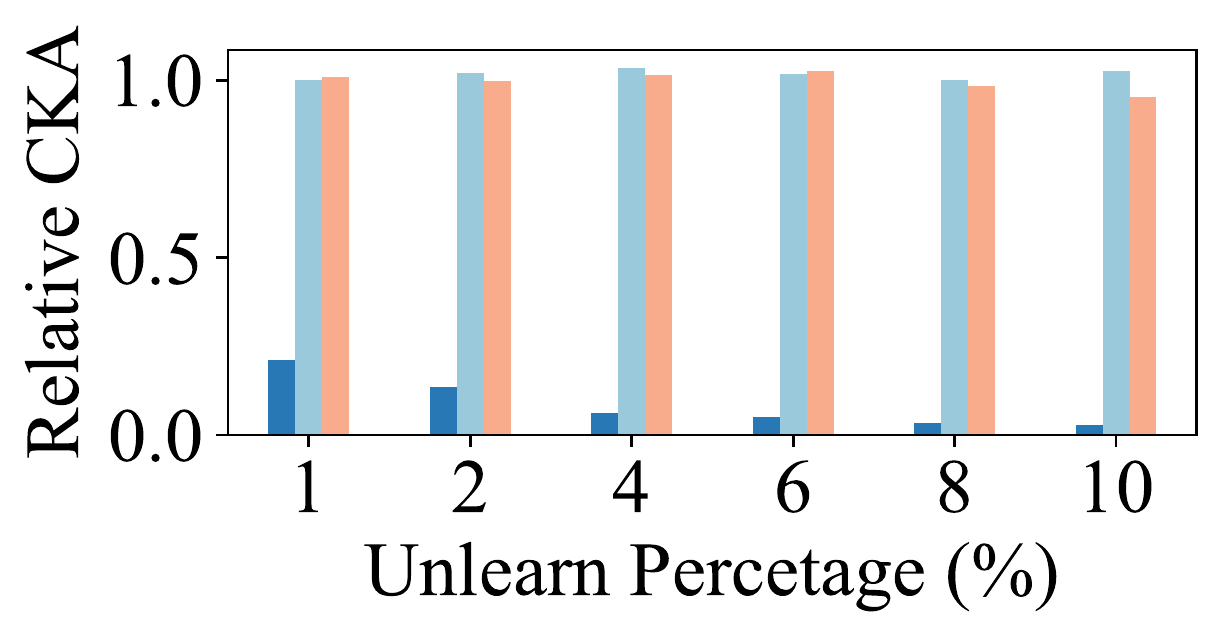}
    }
    \subfigure[LGN]{
        \includegraphics[width=0.47\linewidth]{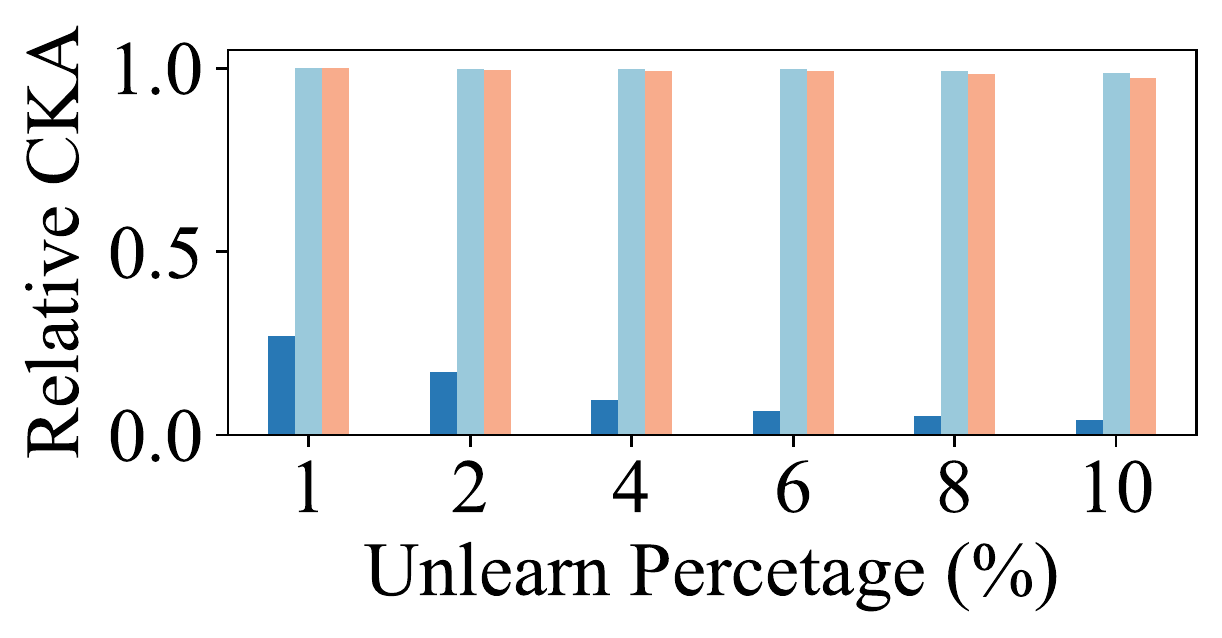}
    }
    \caption{The correspondence, i.e., relative CKA, of different parameters before and after unlearning, where UE (unlearn) and UE (remain) denote User Embedding (UE) of unlearned target users and remaining users respectively.
    We unlearn, i.e., retrain from scratch, up to 10\% of users.}
    \label{fig:cka}
\end{figure}

Inspired by the selective forgetting via reverse Newton update~\citep{sekhari2021remember,golatkar2020eternal}, AU methods can directly unlearn a data point $z$ by setting $\epsilon = -1$ as follow:
\begin{equation}
    \theta^*_{\lnot z} = \theta^* - \mathcal{I}(z).
\end{equation}
Note that this one-step update based on IF will not affect the original learning workflow, which means it can be directly applied to a pre-existing model.

\subsubsection{User Selection}
Existing AU methods cannot fully meet the needs of recommendation tasks.
Although AU methods can theoretically enhance unlearning efficiency to a large extent, the computational overhead of Equation~\ref{equ:if} is still prohibitive, especially for large-scale recommendation models in practice.
The latest AU methods approximate the calculation to enhance efficiency~\citep{wu2022puma,mehta2022deep}, which reduces calculation accuracy.

To better understand the influence of target data, we follow the definition of strong unlearning and compare the parameter divergence between the original recommendation model and the unlearned one, i.e., before and after unlearning.
Specifically, we perform an empirical study on a benchmark recommendation dataset, i.e., MovieLens 1M, and report the results in Figure~\ref{fig:cka}.
The unlearned recommendation models are obtained by retraining from scratch.
As for the unlearning target, we perform user-wise unlearning, i.e., unlearning all data of a target user, because sample-wise unlearning can only bring insignificant parameter changes.
As for the divergence metric, we use relative Centered Kernel Alignment (CKA)~\citep{kornblith2019similarity} to measure the correspondences between the model parameters from different initialization.
The larger the CKA value, the smaller the divergence.
To compare CKA across different parameters, we scale CKA by the average CKA of the original models, obtaining the relative CKA. 
Formally, the relative CKA between the original model $\theta$ and the unlearned model $\theta_\lnot$ is computes as:
\begin{equation}
    \text{Relative CKA}(\theta, \theta_\lnot) = \frac{\text{CKA}(\theta, \theta_\lnot)}{\sum_{i, j \in \mathcal{P}_2([M])}\text{CKA}(\theta_i, \theta_j) / C_M^2},
\end{equation}
where $\mathcal{P}_2([M])$ and $C_M^2$ represent the set of combinations and the number of combinations, respectively, for selecting two objects from a total of M objects.
We set $M = 10$ in our empirical study.

We observe from Figure~\ref{fig:cka} that user-wise unlearning mainly influences the embedding of the target users, i.e., UE (unlearn), leaving the embedding of the remaining users and other parameters barely affected.
This observation enlightens us that computing the IF of all model parameters is not necessary.
To this end, rather than updating all model parameters, we propose to only compute the IF of the target users and update their user embedding selectively.
Through this approach, we can effectively decrease the computational overhead at a fundamental level without compromising the accuracy of the calculations.

\begin{figure}[t]
    \centering
    \includegraphics[width=0.7\linewidth]{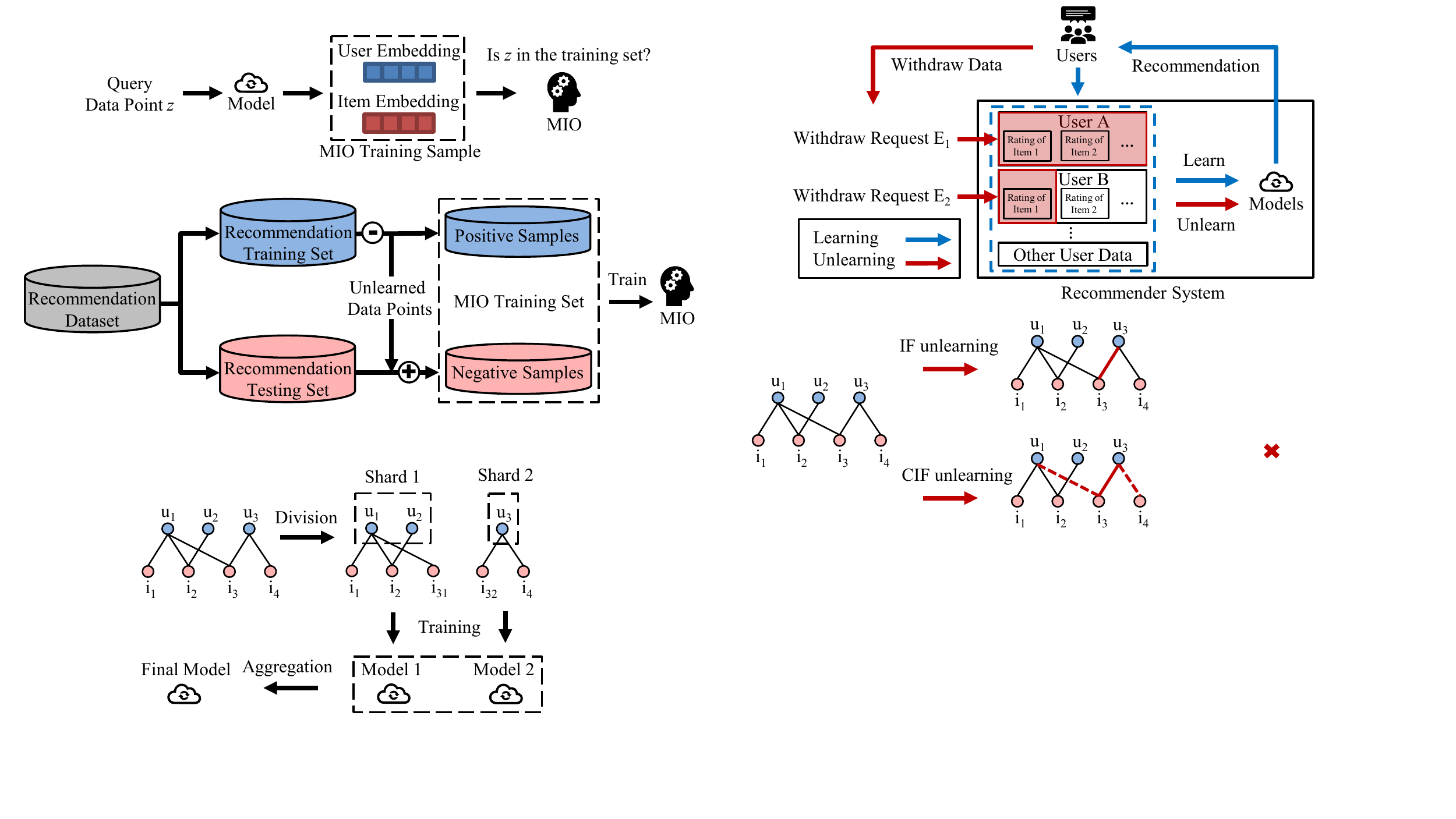}
    \caption{Difference between IF unlearning and CIF unlearning where rating $(u_3, i_3)$ is unlearned. 
    Both IF and CIF do not change the original learning workflow. 
    IF only focuses on the influence of the target rating, i.e., the red solid edge $(u_3, i_3)$, without considering associated ratings. 
    To preserve the collaboration of remaining users and items, CIF computes the collaborative influence by further considering associated ratings, i.e., the red dotted edges $(u_1, i_3)$ and $(u_3, i_4)$.
    }
    \label{fig:cif}
\end{figure}

\subsubsection{Collaboration Preservation}

The IF-based unlearning method regards each data point as an unassociated participator that contributes to the model \textit{individually}.
However, in the context of recommendation, all data points contribute to the model \textit{collaboratively}.
Unlearning the lineage of one data point can influence the lineage of others.
Thus, we further incorporate a collaborative component into influence function to boost model utility (P3).
Figure~\ref{fig:cif} illustrates the difference between IF and Collaborative Influence Function (CIF).

Instead of removing the target data from the training set, the basic idea of CIF is to replace the target data with remaining collaborative information.
Thus, unlearning a data point $z$ is now described as:
\begin{equation}
    \theta^*_{z\to \bar{z}} = \underset{\theta}{\operatorname{\arg\min}}\hspace{1mm}\sum_{i=1}^n \ell(z_i, \theta) - \ell(z, \theta) + \ell(\bar{z}, \theta),
\end{equation}
where we add the last term as a \textit{collaborative component}.
In the RS, a user interacts with a number of items and vice versa.
The collaborative component is designed to restore collaboration across the remaining users and items.
In this paper, we use the average rating of the user in the target data point to represent $\bar{z}$. 
If there was no remaining rating of the user, we use the average rating of the item instead.

Then following the derivation of IF, we define the $\epsilon$-weighted parameter as
\begin{equation}
    \theta^*_{\epsilon, z\to \bar{z}} = \arg\min_\theta\sum_{i=1}^n\ell(z_i, \theta) - \epsilon[\ell(z, \theta) - \ell(\bar{z}, \theta)].
\end{equation}
Thus, the CIF is given by
\begin{equation}
    \mathcal{I}_c(z) = H^{-1}_{\theta^*}[\ell(z, \theta^*) - \ell(\bar{z}, \theta^*)].
\end{equation}
Please refer to~\ref{sec:cif} for the details of CIF derivation.
Based on CIF, we can collaboratively unlearn a data point $z$ by setting $\epsilon = 1$ as follow:
\begin{equation}
    \theta^*_{z\to\bar{z}} = \theta^* + \mathcal{I}_c(z).
\end{equation}
In this way, our proposed CIF can not only achieve direct unlearning, but also restore collaboration for recommendation (P2).

\subsubsection{Computation of SCIF}
With the help of user selection, we greatly reduce the computational overhead of IF.
There are still two challenges to compute the influence on the target user embedding.
Firstly, it is relatively expensive to compute IF and CIF, because it involves computing the inverse of the Hessian matrix.
Following~\cite{koh2017understanding,basu2020second}, we compute Hessian-Vector Product (HVP) to reduce computational overhead in our experiments.
Specifically, we use conjugate gradients~\citep{shewchuk1994introduction} which transforms the HVP computation into a second-order optimization problem.
In this way, the second-order influence can be computed at a similar cost to that of the first-order gradient.

Secondly, the assumption of finding the minimizer $\theta^*$ is hard to hold in practice.
Machine learning models, including recommendation models, are often trapped in a local minimizer $\tilde{\theta}$ instead of the global minimizer $\theta^*$.
In order to approximate the influence more precisely, we compute CIF of $z$ as follows:
\begin{equation}
    \mathcal{I}_c(z) = H^{-1}_{\tilde{\theta}}[\nabla_\theta(z, \tilde{\theta}) - \ell(\bar{z}, \tilde{\theta}) + g],
\end{equation}
where $g = \sum_{i=1}^n\nabla_\theta\ell(z_i, \tilde{\theta})$ is the compensation term for the gap between local minimizer and global minimizer.
%

\section{Completeness Evaluation}\label{sec:eva}

As we mentioned in Section~\ref{sec:goal}, completely unlearning the associated data lineage is the most fundamental requirement of unlearning (P1).
To the best of our knowledge, there is no specific metric to evaluate the \textit{level} of unlearning completeness in recommendation tasks.
Existing work mainly focuses on the unlearning problem in classification tasks~\citep{schelter2021hedgecut,baumhauer2020machine,bourtoule2021machine}, where it is easy to judge the completeness by analyzing the logit output, which refers to the re-scaled logarithmic probabilities indicating the likelihood of the data point belonging to particular classes.
However, in other tasks, e.g., regression and recommendation, completeness is difficult to measure.
What is more, the above evaluation method is only designed for weak unlearning, i.e., analyzing the output of models, but not for strong unlearning, which makes the evaluation more challenging. 
For example, no matter whether the model parameters change or not, existing evaluation methods consider the model as completely unlearned, if it outputs a uniform logit for the unlearning target.
Hence, we aim at proposing a general metric to evaluate the level of completeness for strong unlearning, 
This metric will differ from the one used for weak unlearning, which is currently employed in existing works focused on classification tasks.

\subsection{Evaluation for Strong Unlearning}

Conceptually, strong unlearning generates an unlearned model that has never seen the target data.
In other words, the target data is in the training set of the original model, while out of that of the unlearned model.
Based on the deduction above, we define completeness of strong unlearning as follow:

\begin{definition}[Completeness of Strong Unlearning]
    Given a model $\mathcal{M}(D)$ that is trained on dataset $D$ and a data point $z$ in $D$, we define $z$ as being completely unlearned if $z$ is not in the training set of $\mathcal{M}_{\lnot z}(D)$.
\end{definition}

In this way, we transform a complex task, i.e., completeness evaluation, into a binary classification task. 
For the remainder of this paper, we use the term \textit{completeness} to refer to the completeness of strong unlearning.

\subsection{Evaluation with MIO}

To determine whether a given data point was in the training set, we use a widely acknowledged attack method, i.e., membership inference attack.
Following this idea, we define an ideal concept, i.e., Membership Inference Oracle (MIO), to evaluate unlearning completeness.

\begin{definition}[Membership Inference Oracle]
    We define a membership inference attacker as oracle if it can achieve an attack accuracy of 100\%.
\end{definition}

In addition to the binary prediction output by MIO, we can further measure the level of completeness by the output probability.
In fact, using probability is more convincing than binary prediction for several reasons. 
Firstly, binary results have the potential to lose valuable information that could be useful in completeness evaluation. 
Secondly, the threshold for binary predictions, i.e., the critical probability of completeness, can vary significantly across different datasets, making it challenging to determine in practice. 
Therefore, relying on probability rather than binary prediction provides a more accurate and nuanced understanding of completeness evaluation.

In the following part of this section, we will introduce the usage of MIO in completeness evaluation and the training method of an approximated MIO.

\subsubsection{MIO Usage}

\begin{figure}[t]
    \centering
    \includegraphics[width=0.95\linewidth]{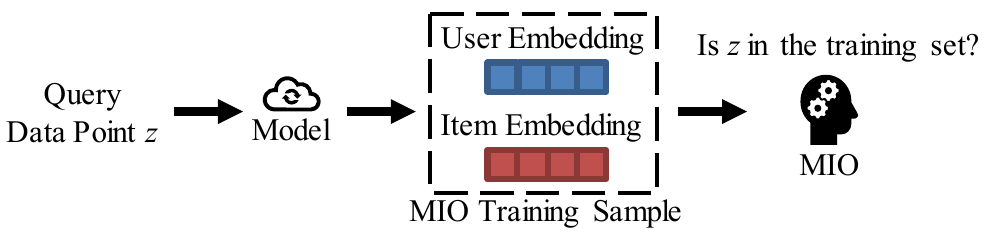}
    \caption{Process of unlearning completeness evaluation by MIO in recommendation tasks.}
    \label{fig:eva}
\end{figure}

As shown in Figure~\ref{fig:eva}, MIO generally follows the normal procedure of membership inference to evaluate unlearning completeness.
The key difference between MIO and other membership inference attacks is the training samples of the attacker.
Existing work uses logit output~\citep{shokri2017membership} (for classification tasks) or training loss~\citep{wu2020characterizing,yu2021does} (for a wider range of tasks) to train the attacker, which means that the training samples are model outputs. 
Therefore, directly applying the above membership inference attacks can only evaluate completeness at the weak-unlearning level.
In contrast, our proposed MIO takes user and item embedding as training samples, which brings two advantages.
Firstly, user and item embedding is a part of model parameters.
Using it as training samples can evaluate completeness at the strong-unlearning level.
Secondly, compared with training loss, user and item embedding reflects the collaborative property of the query data point, which is the key insight of collaborative filtering. 
For user-wise unlearning, we take user embedding and the average item embedding of interacted items as training samples.
As investigating the choice of attackers is not the focus of this paper, we use neural networks, a powerful algorithm that has been successfully applied in numerous fields, to implement an approximated MIO in our experiments.

\subsubsection{Approximated MIO Training}

\begin{figure}[t]
    \centering
    \includegraphics[width=\linewidth]{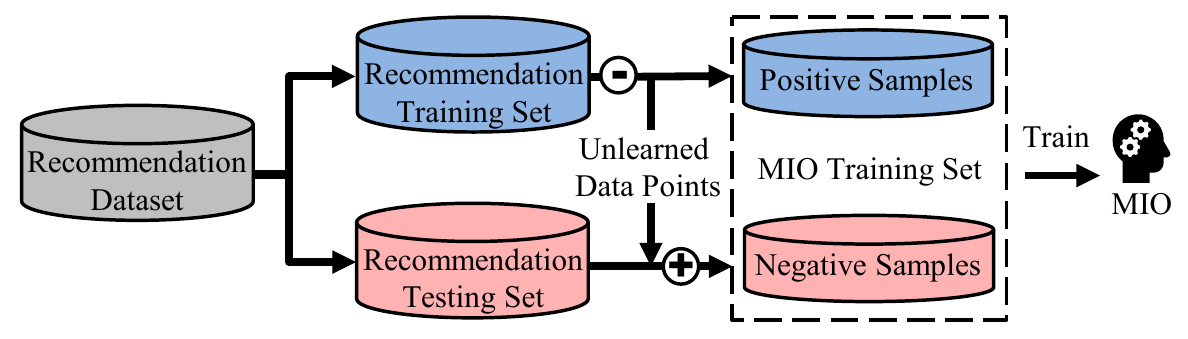}
    \caption{MIO training workflow in recommendation tasks.}
    \label{fig:train}
\end{figure}

In order to train an approximated MIO, we construct a balanced training set that consists of both positive and negative samples.
As shown in Figure~\ref{fig:train}, the negative samples are the combination of unlearned data points and testing data while positive samples are the remaining training data.
In the context of our completeness evaluation setting, MIO serves as a strong white-box attacker owing to its ability to gain access to both the original dataset and the model parameters.
As such, there arises no necessity for conducting shadow training.
In other words, we can adopt the view that the original recommendation dataset functions as the shadow dataset, while the trained model serves as the shadow model.
Please refer to \cite{salem2018ml} for more details of membership inference attacks.

\section{Experiments}

To comprehensively evaluate the effectiveness of our proposed method, we conduct experiments on two widely used datasets in terms of three principles, i.e., unlearning completeness, unlearning efficiency, and model utility. 
In addition, we also perform an ablation study to gain a better understanding of our proposed method.

\subsection{Dataset}

We evaluate our framework on two publicly accessible datasets: MovieLens 1M (ML1M)\footnote{https://grouplens.org/datasets/movielens/} and Amazon Digital Music (ADM)\footnote{http://jmcauley.ucsd.edu/data/amazon/}. 
These two datasets are widely used as benchmarks to evaluate recommendation models~\citep{harper2015movielens,he2016ups,rappaz2021recommendation}. 
%
%
%
%
We filter out the users and items that have less than 5 interactions. 
In order to construct a balanced training set for MIO, half of the ratings are used for training and the rest for testing.
Table~\ref{tab:dataset} summarizes the statistics of the above two datasets.

\begin{table}
\caption{Summary of datasets.}
\label{tab:dataset}
\begin{tabular}{lrrrr}  
\toprule
Dataset & User \#   & Item \#   & Rating \# & Sparsity \\
\midrule
ML1M    & 6,040     & 3,706     & 1,000,209 & 95.532\%\\
ADM     & 478,235   & 266,414   & 836,006   & 99.999\%\\
\bottomrule
\end{tabular}
\end{table}

\subsection{Recommendation Models}

The basic idea of modern recommendation models is Collaborative Filtering (CF)~\citep{hu2008collaborative,shi2014collaborative}.
We evaluate different unlearning methods on two representative deep CF models.
\begin{itemize}
    \item \textbf{NMF} Neural Matrix Factorization (NMF)~\citep{he2017neural} is a well-recognized CF model based on matrix factorization.
    \item \textbf{LGN} LightGCN (LGN) is the State-Of-The-Art (SOTA) CF model~\citep{he2020lightgcn}, which simplifies graph convolution networks to improve recommendation performance.
\end{itemize}

\subsection{Compared Methods}

We select three representative unlearning methods for comparison.
\begin{itemize}
    \item \textbf{Retrain} Retraining from scratch is a straightforward unlearning method that achieves strong unlearning with heavy computational overhead. 
    \item \textbf{RecEraser} This is the SOTA recommendation unlearning method, which follows the divide-aggregate framework~\citep{chen2022recommendation}.
    Specifically, we divide the dataset into 8 shards for RecEraser in this paper.
    \item \textbf{IF} We also compare with the unlearning method based on original IF~\citep{sekhari2021remember}.
\end{itemize}

\begin{table}
\caption{Results of unlearning completeness. Note that IF and SCIF do not interfere with the learning process. Thus, they share the same result with Retrain during learning.}
\label{tab:comp}
\centering
\resizebox{\linewidth}{!}{
\begin{tabular}{lc|rr|rr}
\toprule
& & \multicolumn{2}{c}{ML1M} & \multicolumn{2}{c}{ADM}\\
& & ACC & AUC & ACC & AUC \\
\midrule
\midrule
\multirow{2}{*}{LGN - learn} & Retrain & 0.741 & 0.787 & 0.756 & 0.792 \\
& RecEraser  & 0.626 & 0.681 & 0.621 & 0.673 \\
\midrule
\multirow{4}{*}{LGN - rand@2.5} & Retrain & 0.552 & 0.571 & 0.547 & 0.583 \\
& RecEraser  & 0.509 & 0.523 & 0.511 & 0.515 \\
& IF    & 0.568 & 0.573 & 0.564 & 0.579 \\
& SCIF   & 0.575 & 0.580 & 0.573 & 0.581 \\
\midrule
\multirow{4}{*}{LGN - rand@5} & Retrain & 0.544 & 0.558 & 0.547 & 0.561 \\
& RecEraser  & 0.499 & 0.506 & 0.492 & 0.509 \\
& IF    & 0.559 & 0.562 & 0.557 & 0.573 \\
& SCIF   & 0.561 & 0.567 & 0.565 & 0.579 \\
\midrule
\midrule
\multirow{2}{*}{NMF - learn} & Retrain & 0.748 & 0.804 & 0.747 & 0.793 \\
& RecEraser  & 0.609 & 0.694 & 0.610 & 0.671 \\
\midrule
\multirow{4}{*}{NMF - rand@2.5} & Retrain & 0.554 & 0.569 & 0.551 & 0.588 \\
& RecEraser  & 0.516 & 0.532 & 0.513 & 0.526 \\
& IF    & 0.576 & 0.579 & 0.569 & 0.582 \\
& SCIF   & 0.585 & 0.589 & 0.581 & 0.590 \\
\midrule
\multirow{4}{*}{NMF - rand@5} & Retrain & 0.549 & 0.565 & 0.556 & 0.571 \\
& RecEraser  & 0.500 & 0.506 & 0.494 & 0.504 \\
& IF    & 0.565 & 0.573 & 0.566 & 0.575 \\
& SCIF   & 0.570 & 0.578 & 0.575 & 0.587 \\
\bottomrule
\end{tabular}
}
\end{table}

\subsection{Parameter Settings}

We tune the hyper-parameters using grid search to obtain the optimal performance.
For model-specific hyper-parameters, we tune them based on the suggestions from their original papers.
The model parameters are initialized with a Gaussian distribution $\mathcal{N}(0, 0.01^2)$.
Specifically, we set the learning rate to 0.001 and the embedding size to 64.
The total number of epochs $T$ is set to 50 for NMF, and 200 for LGN.
%
%
We ran all models for 10 times and report the average results. 

\begin{figure}[t]
    \centering
    \includegraphics[width=0.9\linewidth]{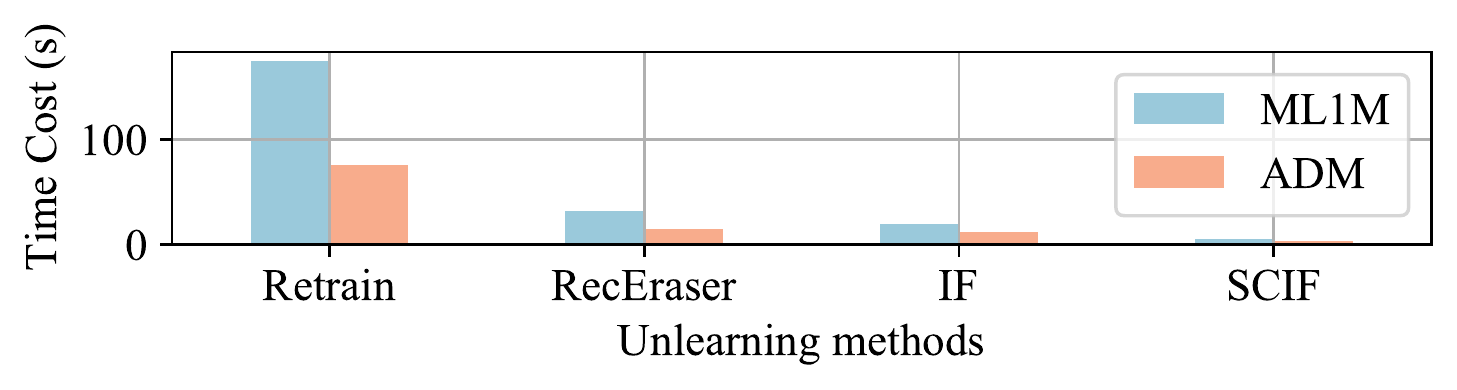}
    \caption{Running time of unlearning on NMF.}
    \label{fig:time}
\end{figure}

\begin{table*}
\caption{Results of recommendation metrics. We highlight the top results except Retrain in bold.}
\label{tab:res}
\resizebox{\linewidth}{!}{
\begin{tabular}{lc|rrrr|rrrr}
\toprule
\multicolumn{10}{c}{MovieLens 1M (ML1M)} \\
\cmidrule{3-10}
& & NDCG@5 & HR@5 & NDCG@10 & HR@10 &  NDCG@15 & HR@15 & NDCG@20 & HR@20 \\
\midrule
\multirow{2}{*}{LGN - learn} & Retrain  & 0.4874 & 0.5231 & 0.5109 & 0.5982 & 0.4838 & 0.6265 & 0.4583 & 0.6404 \\
& RecEraser    & 0.3139 & 0.3368 & 0.3359 & 0.3962 & 0.3125 & 0.4042 & 0.3001 & 0.4183\\
\midrule
\multirow{4}{*}{LGN - rand@2.5} & Retrain  & 0.4873 & 0.5221 & 0.5107 & 0.5977 & 0.4847 & 0.6259 & 0.4582 & 0.6758\\
& RecEraser    & 0.3061 & 0.3442 & 0.3183 & 0.3827 & 0.3045 & 0.4227 & 0.2997 & 0.4293\\
& IF      & 0.4141 & 0.4427 & 0.4391 & 0.5178 & 0.4104 & 0.5408 & 0.3983 & 0.5592\\
& SCIF     & \textbf{0.4507} & \textbf{0.4789} & \textbf{0.4662} & \textbf{0.5564} & \textbf{0.4413} & \textbf{0.5884} & \textbf{0.4127} & \textbf{0.5928}\\
\midrule
\multirow{4}{*}{LGN - rand@5} & Retrain  & 0.4861 & 0.5227 & 0.5084 & 0.5951 & 0.4852 & 0.6259 & 0.4573 & 0.6409\\
& RecEraser    & 0.3285 & 0.3304 & 0.3403 & 0.3805 & 0.3287 & 0.4046 & 0.3147 & 0.4175\\
& IF      & 0.4187 & 0.4271 & 0.4258 & 0.4953 & 0.4168 & 0.5257 & 0.3897 & 0.5251\\
& SCIF     & \textbf{0.4457} & \textbf{0.4797} & \textbf{0.4733} & \textbf{0.5526} & \textbf{0.4520} & \textbf{0.5817} & \textbf{0.4128} & \textbf{0.5873}\\
\midrule
\multirow{2}{*}{NMF - learn} & Retrain  & 0.5083 & 0.5441 & 0.5266 & 0.6187 & 0.4965 & 0.6376 & 0.4582 & 0.6405 \\
& RecEraser    & 0.3306 & 0.3612 & 0.3507 & 0.4046 & 0.3193 & 0.4210 & 0.3020 & 0.4300\\
\midrule
\multirow{4}{*}{NMF - rand@2.5} & Retrain  & 0.5076 & 0.5427 & 0.5270 & 0.6186 & 0.4957 & 0.6370 & 0.4578 & 0.6402\\
& RecEraser    & 0.3194 & 0.3615 & 0.3367 & 0.4160 & 0.3252 & 0.4233 & 0.2940 & 0.4213\\
& IF      & 0.4319 & 0.4764 & 0.4582 & 0.5286 & 0.4337 & 0.5538 & 0.4032 & 0.5520\\
& SCIF     & \textbf{0.4603} & \textbf{0.4995} & \textbf{0.4780} & \textbf{0.5794} & \textbf{0.4533} & \textbf{0.5990} & \textbf{0.4312} & \textbf{0.5877}\\
\midrule
\multirow{4}{*}{NMF - rand@5} & Retrain  & 0.5055 & 0.5419 & 0.5256 & 0.6174 & 0.4949 & 0.6360 & 0.4580 & 0.6406\\
& RecEraser    & 0.3576 & 0.3464 & 0.3581 & 0.3913 & 0.3472 & 0.4164 & 0.3193 & 0.4099\\
& IF      & 0.4341 & 0.4486 & 0.4487 & 0.5243 & 0.4158 & 0.5400 & 0.3840 & 0.5317\\
& SCIF     & \textbf{0.4616} & \textbf{0.5044} & \textbf{0.4890} & \textbf{0.5729} & \textbf{0.4601} & \textbf{0.5813} & \textbf{0.4257} & \textbf{0.5866}\\
\midrule
\midrule
\multicolumn{10}{c}{Amazon Digital Music (ADM)}\\
\cmidrule{3-10}
& & NDCG@5 & HR@5 & NDCG@10 & HR@10 &  NDCG@15 & HR@15 & NDCG@20 & HR@20 \\
\midrule
\multirow{2}{*}{LGN - learn} & Retrain  & 0.5743 & 0.7519 & 0.4603 & 0.7931 & 0.3828 & 0.8017 & 0.3318 & 0.8055 \\
& RecEraser    & 0.3745 & 0.4921 & 0.3055 & 0.5184 & 0.2610 & 0.5270 & 0.2148 & 0.5347\\
\midrule
\multirow{4}{*}{LGN - rand@2.5} & Retrain  & 0.5742 & 0.7518 & 0.4594 & 0.7945 & 0.3824 & 0.8017 & 0.8054 & 0.8054 \\
& RecEraser    & 0.3665 & 0.4947 & 0.3127 & 0.5263 & 0.2489 & 0.5377 & 0.2251 & 0.5282\\
& IF      & 0.5011 & 0.6557 & 0.3963 & 0.6851 & 0.3275 & 0.6898 & 0.2906 & 0.7014 \\
& SCIF     & \textbf{0.5258} & \textbf{0.6907} & \textbf{0.4296} & \textbf{0.7393} & \textbf{0.3586} & \textbf{0.7474} & \textbf{0.3096} & \textbf{0.7386} \\
\midrule
\multirow{4}{*}{LGN - rand@5} & Retrain  & 0.5746 & 0.7529 & 0.4595 & 0.7936 & 0.3824 & 0.8017 & 0.3317 & 0.8054 \\
& RecEraser    & 0.3934 & 0.4817 & 0.2997 & 0.5321 & 0.2697 & 0.5143 & 0.2302 & 0.5258 \\
& IF      & 0.4839 & 0.6171 & 0.4043 & 0.6781 & 0.3182 & 0.6671 & 0.2819 & 0.6708\\
& SCIF     & \textbf{0.5351} & \textbf{0.6993} & \textbf{0.4296} & \textbf{0.7245} & \textbf{0.3593} & \textbf{0.7383} & \textbf{0.2991} & \textbf{0.7449} \\
\midrule
\multirow{2}{*}{NMF - learn} & Retrain  & 0.5606 & 0.7428 & 0.4549 & 0.7914 & 0.3822 & 0.8017 & 0.3318 & 0.8057 \\
& RecEraser    & 0.3620 & 0.4840 & 0.3015 & 0.5175 & 0.2525 & 0.5319 & 0.2316 & 0.5293 \\
\midrule
\multirow{4}{*}{NMF - rand@2.5} & Retrain  & 0.5615 & 0.7427 & 0.4551 & 0.7915 & 0.3819 & 0.8018 & 0.3317 & 0.8057\\
& RecEraser    & 0.3591 & 0.4838 & 0.2999 & 0.5246 & 0.2509 & 0.5259 & 0.2094 & 0.5291 \\
& IF      & 0.4724 & 0.6510 & 0.4021 & 0.6959 & 0.3388 & 0.6894 & 0.2811 & 0.7006 \\
& SCIF     & \textbf{0.5141} & \textbf{0.6791} & \textbf{0.4263} & \textbf{0.7273} & \textbf{0.3468} & \textbf{0.7333} & \textbf{0.3028} & \textbf{0.7504} \\
\midrule
\multirow{4}{*}{NMF - rand@5} & Retrain  & 0.5618 & 0.7434 & 0.4554 & 0.7916 & 0.3816 & 0.8016 & 0.3319 & 0.8058\\
& RecEraser    & 0.3955 & 0.4707 & 0.3112 & 0.5327 & 0.2663 & 0.5177 & 0.2282 & 0.5118 \\
& IF      & 0.4747 & 0.6240 & 0.3973 & 0.6772 & 0.3286 & 0.6691 & 0.2847 & 0.6639 \\
& SCIF     & \textbf{0.5114} & \textbf{0.6956} & \textbf{0.4252} & \textbf{0.7288} & \textbf{0.3498} & \textbf{0.7458} & \textbf{0.3028} & \textbf{0.7440} \\
\bottomrule
\end{tabular}
}
\end{table*}

To study the effect of unlearning, we define a random user-wise withdrawal request, i.e., rand@$\alpha$, which denotes randomly unlearning $\alpha\%$ of users. 
We vary $\alpha$ in \{2.5, 5\} for both datasets.

\subsubsection{Unlearning Completeness}

We implement an approximated MIO via a basic three-layer (64, 16, 4) neural network with ReLu and Softmax as activation functions for hidden layers and the output layer respectively.
Specifically, we train the MIO via stochastic gradient descent with 100 epochs and a learning rate of 0.001.
The MIO outputs the probability of the queried data point being in the training set. %
To evaluate the unlearning completeness, we query MIO with the unlearned data points.
Ideally, MIO outputs 1 (being in the training set) for the original model while outputs 0 (not being in the training set) for the unlearned model.
In addition to Accuracy (ACC), we also use Area Under the ROC Curve (AUC)~\citep{fawcett2006introduction} to better interpret the output probability.
We set the threshold as 0.5 for ACC.
AUC $=1$ indicates a maximum performance while AUC $=0.5$ indicates a performance equivalent to random guessing~\citep{salem2018ml,backes2017walk2friends,jia2019memguard}. 
We report ACC and AUC of the unlearned data points in Table~\ref{tab:comp}.
From it, we have the following observations:
\begin{itemize}
    \item For Retrain, i.e., ground truth, MIO achieves all ACCs over 0.74 and AUCs over 0.78 in the learning stage while all ACCs below 0.56 and AUCs below 0.59 in the unlearning state, which means our proposed approximated MIO can effectively distinguish the unlearned data points with the original ones.
    \item In the learning workflow, RecEraser can only achieve ACCs and AUCs below 0.70.
    This gap between Retrain and RecEraser indicates that the divide-aggregate methods such as RecEraser decrease model utility.
    \item IF and SCIF achieve similar performance in terms of completeness.
    As they do not interfere with the learning workflow, they share the same performance with Retrain during learning.
    In the unlearning stage, they achieve ACCs below 0.58 and AUCs below 0.60.
    Although the performances of IF and SCIF appear to be slightly worse than Retrain, they are generally able to achieve comparable levels of unlearning completeness.
    The observations above are consistent for both NMF and LGN across two datasets.
\end{itemize}

\subsubsection{Unlearning Efficiency}\label{sec:eff}

In this paper, we aim at accelerating RS's response speed to withdrawal requests.
Therefore, we focus on measuring time efficiency.
Specifically, we randomly unlearn 5\% of users and report the corresponding unlearning time cost.
Note that we only report the running time of unlearning, because (i) our focus in this paper is unlearning in the RS, and (ii) all compared methods cost similar running time during learning.

We report the unlearning time of NMF in Figure~\ref{fig:time} for conciseness. 
From it, we observe that RecEraser and IF enhance the unlearning efficiency to a certain extent, but it is not comparable to SCIF.
To be specific, considering Retrain as a baseline, RecEraser and IF improve the time efficiency by 5.2 and 7.7 times respectively, while SCIF improves the efficiency by 28.5 times.

\subsubsection{Model Utility}

We use two widely used metrics, i.e., Precision (Prec), Recall, Normalized Discounted Cumulative Gain (NDCG), and Hit Ratio (HR)~\citep{he2015trirank,xue2017deep,ji2020dual}, to comprehensively evaluate the recommendation performance.
We examine the ranked list at top-$K (K \in \{5, 10, 15, 20\})$ for all metrics and report the results during both learning and unlearning workflows.
We report the results of all metrics in Table~\ref{tab:res}. 
From it, we obtain the following observation:
\begin{itemize}
    \item As the ground truth method, Retrain achieves the best results throughout all recommendation metrics at the expense of heavy computational overhead.
    Note that during the learning stage, IF and SCIF have the same performance as Retrain, since they do not change the original learning workflow.
    \item RecEraser is a divide-aggregate method that is devised for recommendation tasks. 
    However, compared with the other three methods, it achieves the worst results in recommendation tasks.
    This is because recommendation is an association-sensitive task.
    CF models, e.g., LGN and NMF, exploit this characteristic by elaborating the collaboration between users and items.
    But divide-aggregate methods, e.g., RecEraser, impair such collaboration in both learning and unlearning workflows, as has been illustrated in Figure~\ref{fig:coll}.
    Our experimental results show that RecEraser causes a noticeable drop in recommendation performance during both learning and unlearning.
    \item In general, IF and SCIF achieve similar performance.
    On the one hand, their results are better than that of RecEraser, which means that, compared with divide-aggregate unlearning methods, they can alleviate the damage to the collaboration in recommendation tasks. 
    On the other hand, their results are worse than that of Retrain, indicating that although IF/SCIF-based methods mitigate the damage, they also impair collaboration to some extent. 
   Moreover, it is remarkable that our proposed SCIF achieves better performance than IF with much less computational overhead.
    \item Our proposed SCIF achieves the second-best results among all the comparison methods.
    There is a consistent improvement of SCIF against IF across all recommendation metrics.
    The key difference w.r.t. recommendation performance between SCIF and IF is that SCIF adds a collaborative component which contains the information on remaining ratings. 
    As a result, this collaborative component contributes to restoring collaboration across remaining users and items.
\end{itemize}


\begin{figure}[t]
    \centering
    \includegraphics[width=0.5\linewidth]{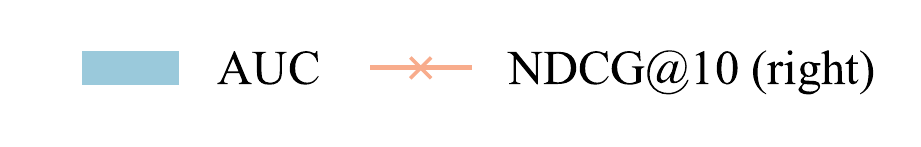}\\
    \subfigure[ML1M - LGN]{
        \includegraphics[width=0.46\linewidth]{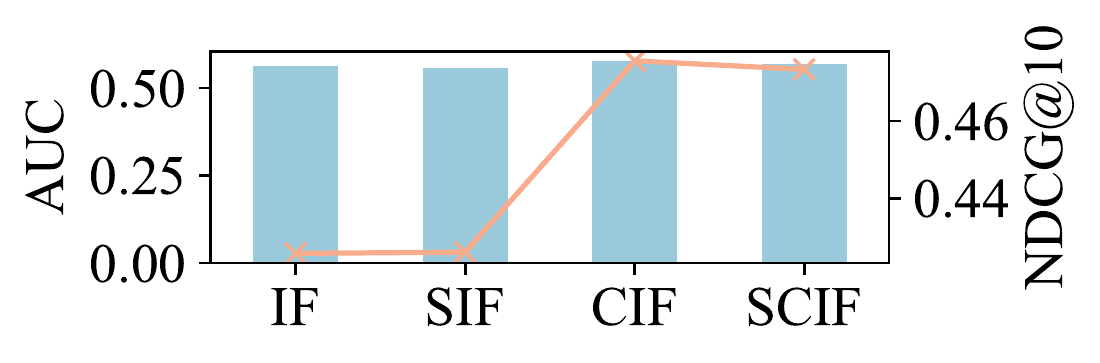}
    }
    \subfigure[ADM - LGN]{
        \includegraphics[width=0.46\linewidth]{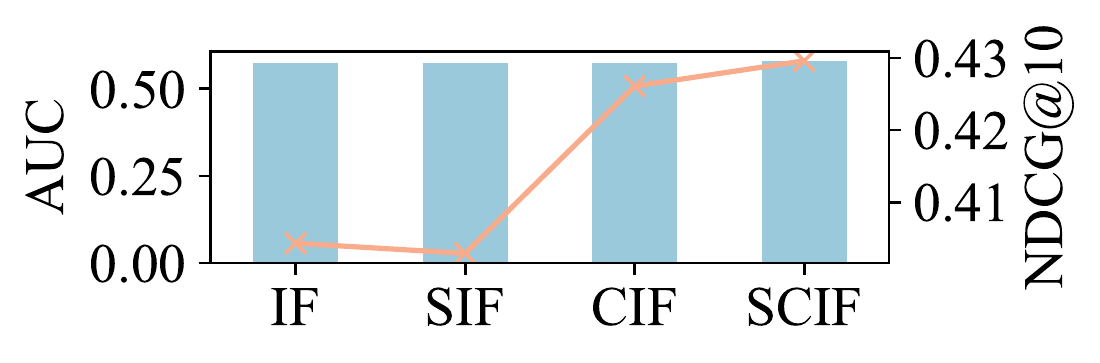}
    }
    \subfigure[ML1M - NMF]{
        \includegraphics[width=0.46\linewidth]{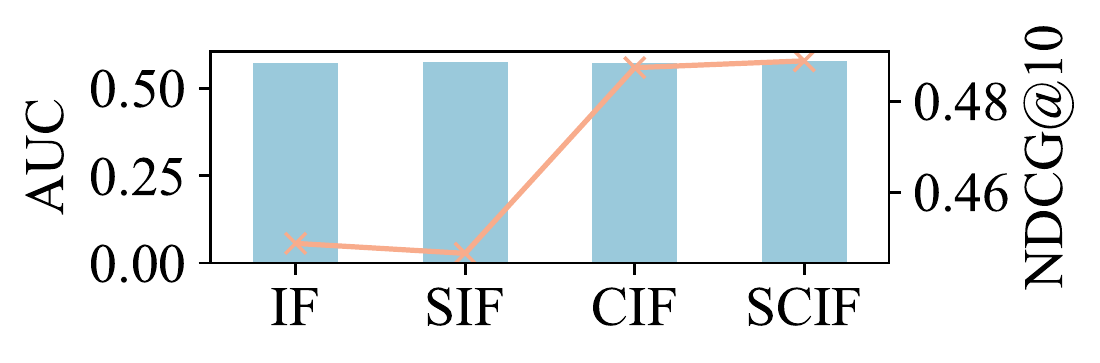}
    }
    \subfigure[ADM - NMF]{
        \includegraphics[width=0.46\linewidth]{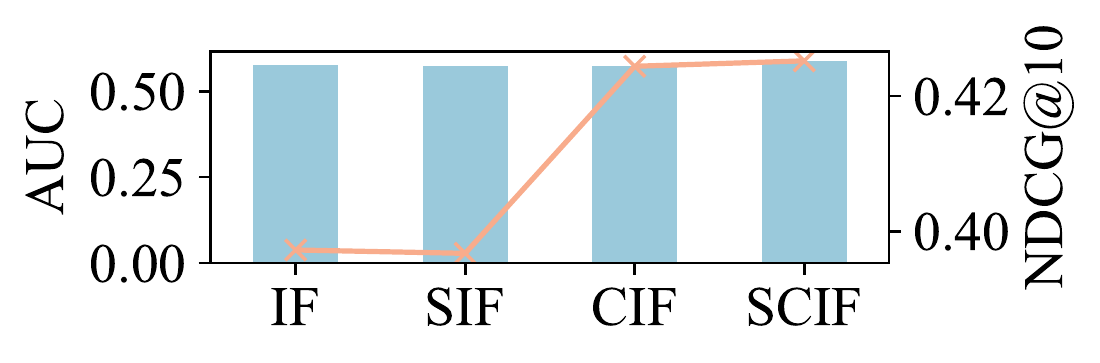}
    }
    \caption{The results of ablation study on user selection, i.e., SIF, and collaboration preservation, i.e., CIF, where we perform rand@5 unlearning.}
    \label{fig:abl}
\end{figure}

\subsection{Ablation Study}

To fully investigate the effectiveness of our proposed method, we conduct an ablation study on \textit{user selection} and \textit{collaboration preservation}.
Specifically, we additionally compare with SIF (IF with user selection only), and CIF (IF with collaboration preservation only) regarding three unlearning principles.

For unlearning completeness (P1) and model utility (P3), we report the results in Figure~\ref{fig:abl}.
From it, we observe that (i) all compared methods achieve an AUC under 0.58, indicating that neither user selection nor collaboration preservation damages P1;
(ii) the methods with user selection, i.e., SIF and SCIF, achieve comparable performances in P3 (NDCG@10) with their original methods, i.e., IF and CIF, indicating that user selection does not have a negative impact on P3;
(iii) the methods with collaboration preservation, i.e., CIF and SCIF, achieve a significant improvement in P3 over their original methods, i.e., IF and SIF, indicating that collaboration reservation enhances P3.

For unlearning efficiency (P2), we have reported the running times of IF and SCIF in Figure~\ref{fig:time} and analyzed them in Section~\ref{sec:eff}.
Since the computational overhead of the collaborative component is almost negligible compared with that of IF, the running time of SIF and CIF is nearly the same as SCIF and IF, respectively.
We omit the running time of SIF and CIF for conciseness.

\section{Conclusions and Future Work}

In this paper, we identify three design principles of unlearning methods, i.e., unlearning completeness, unlearning efficiency, and model utility, and proposed an extra-efficient recommendation unlearning method based on SCIF to achieve practical data lineage unlearning in RS.
SCIF not only achieves high efficiency in terms of both time and space by directly taking selective parameter updates for the target data, but also enhances recommendation performance by preserving collaboration.
Moreover, we introduce the concept of MIO to evaluate the level of unlearning completeness and used Neural Networks (NN) to implement it in this paper.
The NN-based MIO clearly distinguishes unlearned data points from learned ones and quantitatively analyzes unlearning completeness in our experiments.
Our empirical study has shown that our SCIF-based unlearning method achieves satisfactory level of unlearning completeness.
More importantly, our SCIF-based unlearning method outperforms the state-of-the-art recommendation unlearning method in terms of various recommendation metrics, which demonstrates the superiority of SCIF-based unlearning method in the context of recommendation.

Our future work will focus on the following two aspects.
Firstly, we will adapt our proposed recommendation unlearning method to more complex models, which including incorporating more sources of user-item interaction data and further enhancing the efficiency of influence computing.
Secondly, we will dig deeper into the collaborative component of CIF, and explore more implementations.

\section*{Declaration of Interests}
The authors declare that they have no known competing financial interests or personal relationships that could have appeared to influence the work reported in this paper.


\appendix

\section{Derivation of CIF}\label{sec:cif}

For Collaborative Influence Function (CIF), unlearning a data point $z$ is now described as:
\begin{equation}
    \theta^*_{z\to \bar{z}} = \underset{\theta}{\operatorname{\arg\min}}\hspace{1mm}\sum_{i=1}^n \ell(z_i, \theta) - \ell(z, \theta) + \ell(\bar{z}, \theta).
\end{equation}
To compute its influence, we define $\mathcal{I}_c(z)$ and the $\epsilon$-weighted parameter as follows:
\begin{align}\label{equ:eps}
    & \mathcal{I}_c(z) := \frac{d\theta^*_{\epsilon, z\to\bar{z}}}{d\epsilon}\Big\vert_{\epsilon=0}, \nonumber\\
    & \theta^*_{\epsilon, z\to \bar{z}} = \underset{\theta}{\operatorname{\arg\min}}\hspace{1mm}\sum_{i=1}^n \ell(z_i, \theta) - \epsilon[\ell(z, \theta) - \ell(\bar{z}, \theta)].
\end{align}
Due to the fact that $\theta^*$ does not depend on $\epsilon$, we have
\begin{equation}
    \frac{d\theta^*_{\epsilon, z\to\bar{z}}}{d\epsilon} = \frac{d\Delta_\epsilon}{d\epsilon},
\end{equation}
where $\Delta = \theta^*_{\epsilon, z\to\bar{z}} - \theta^*$ denotes parameter change.
Since $\theta^*_{\epsilon, z\to \bar{z}}$ is a minimizer of (\ref{equ:eps}), we have
\begin{equation}
    \sum_{i=1}^n \nabla\ell(z_i, \theta^*_{\epsilon, z\to\bar{z}}) - \epsilon[ \nabla\ell(z, \theta^*_{\epsilon, z\to\bar{z}}) - \nabla\ell(\bar{z}, \theta^*_{\epsilon, z\to\bar{z}})] = 0.
\end{equation}
Performing a second order Taylor expansion at $\theta^*_{\epsilon, z\to\bar{z}}\to\theta^*$ ($\epsilon\to 0$), we have
\begin{align}\label{equ:exa}
    & \Big[\sum_{i=1}^n \nabla\ell(z_i, \theta^*) - \epsilon[\nabla\ell(z, \theta^*) - \nabla\ell(\bar{z}, \theta^*)]\Big] + \nonumber\\ 
    & \Big[\sum_{i=1}^n \nabla^2\ell(z_i, \theta^*) - \epsilon[\nabla^2\ell(z, \theta^*) - \nabla^2\ell(\bar{z}, \theta^*)]\Big]\Delta_\epsilon\approx 0.
\end{align}
Rearranging (\ref{equ:exa}), we have
\begin{align}
    \Delta_\epsilon\approx & - \Big[\sum_{i=1}^n \nabla^2\ell(z_i, \theta^*) - \epsilon[\nabla^2\ell(z, \theta^*) - \nabla^2\ell(\bar{z}, \theta^*)]\Big]^{-1} \nonumber\\
    & \Big[\sum_{i=1}^n \nabla\ell(z_i, \theta^*) - \epsilon[\nabla\ell(z, \theta^*) - \nabla\ell(\bar{z}, \theta^*)]\Big].
\end{align}
Since $\theta^*$ minimizes $\sum_{i=1}^n \ell(z_i, \theta)$, we have $\sum_{i=1}^n \nabla\ell(z_i, \theta^*) = 0$.
Dropping $o(\epsilon)$ terms, we have
\begin{equation}
    \mathcal{I}_c(z) = H^{-1}_{\theta^*}[\ell(z, \theta^*) - \ell(\bar{z}, \theta^*)],
\end{equation}
where $H_{\theta^*} = \sum_{i=1}^n \nabla^2\ell(z_i, \theta^*)$ denotes the Hessian matrix.


\end{document}